# Research and Development for a Gadolinium Doped Water Cherenkov Detector

Andrew Renshaw[*] for The Super-Kamiokande Collaboration

*Department of Physics and Astronomy, University of California, Irvine, Irvine, CA 92697-4575, USA*

**Abstract**

The proposed introduction of a soluble gadolinium (Gd) compound into water Cherenkov detectors can result in a high efficiency for the detection of free neutrons capturing on the Gd. The delayed 8 MeV gamma cascades produced by these captures, in coincidence with a prompt positron signal, serve to uniquely identify electron antineutrinos interacting via inverse beta decay. Such coincidence detection can reduce backgrounds, allowing a large Gd-enhanced water Cherenkov detector to make the first observation of supernova relic neutrinos and high precision measurements of Japan's reactor antineutrino flux, while still allowing for all current physics studies to be continued. Now, a dedicated Gd test facility is operating in the Kamioka Mine. This new facility houses everything needed to successfully operate a Gd doped water Cherenkov detector. Successful running of this facility will demonstrate that adding Gd salt to SK is both safe for the detector and is capable of delivering the expected physics benefits.

## 1. Introduction

From the original evidence for neutrino oscillations [1], to setting the best flux limit on supernova relic neutrinos (SRN) [2], current multi-kiloton scale water Cherenkov detectors, such as Super-Kamiokande (SK) [3], have provided many clues in the experimental understanding of the neutrino. SK alone has successfully observed solar [4], atmospheric [1], and accelerator [5] neutrinos. However, in spite of the large livetime of the experiment, some analyses are still limited by statistical uncertainty, and would benefit from increasing exposure. Other analyses suffer from background contamination, as in the case of the SRN search, and would benefit more from the development of new background suppression techniques. The proposed introduction of gadolinium (Gd) ions, at 0.1% loading, into a detector such as SK, results in higher than 90% efficiency of free neutrons capturing on the Gd, giving a handle on antineutrinos acting via inverse beta decay, and possibly a method of background reduction for other studies [6]. However, this novel method of doping a water Cherenkov detector with a Gd salt poses many new and interesting challenges. These challenges are currently under intense investigation in both the US and Japan, the cumulative effort now being focused at a multi-million dollar test facility called EGADS (Evaluating Gadolinium's Action on Detector Systems), which is up and running in the Kamioka Mine in Japan, the same location as SK. This paper will cover the current research and development being done at these sites, working towards a large, Gd doped water Cherenkov detector.

[*] *Email address*: arenshaw@uci.edu.

**2. Current Low Energy Physics At Large Water Cherenkov Detectors**

When studying low energy neutrinos at large water Cherenkov detectors, event patterns appear as single Cherenkov rings on the detector wall, coming from a resultant charged particle, most commonly an electron or positron. At the lowest energies radioactivity and spallation products can dominate and mimic the low energy electron (positron) signal, leaving analyses limited by these backgrounds. But due to a strong forward peaking of the neutrino-electron elastic scattering cross section, a solar neutrino signal can be extracted from the data using the directionality of the electrons [4]. However, in the case of the SRN search, positrons resulting from inverse beta decay have very weak directionality associated with the incoming antineutrino (which do not point back at any particular star anyways), while the resulting free neutron will capture on a hydrogen nucleus, releasing a 2.2 MeV gamma. In SK, this gamma results in only ~7 photo-electrons, and thus only detectable with ~20% efficiency. The SK collaboration recently updated their SRN search [7], including new background modeling, an unbinned maximum likelihood method of signal extraction, and the use of data from the first three phases of SK. When the results are compared to the Ando/Sato/Totani SRN model [8], only a one sigma significance, over background, can be seen, while ten years of Gd-doped SK data should discover a signal at better than 3 sigma significance.

**3. GADZOOKS! (Gadolinium Antineutrino Detector Zealously Outperforming Old Kamiokande, Super!)**

*3.1. The Original Idea*

The idea of adding a soluble Gd compound to a large water Cherenkov detector was first proposed by John Beacom and Mark Vagins, back in 2004 [6], in order to see antineutrinos interacting via inverse beta decay. By adding a Gd compound, such that the concentration of free Gd ions is ~0.1%, to the ultrapure water of a detector such as SK, free neutrons inside the tank would have a better than 90% capture efficiency on the Gd nucleus. When this capture occurs, an 8 MeV gamma cascade is produced, giving more than enough light to pinpoint the capture vertex. In the case of inverse beta decay, this capture vertex will have temporal (~ 30 μs) and spatial (< 200 cm) coincidence with the positron coming from the initial interaction, giving a nice "tag" for the event. This will most directly benefit the SRN search, in which the antineutrinos coming from all past supernovae dominantly interact inside the SK detector via inverse beta decay. However, the ability to very efficiently detect free neutrons inside the SK tank opens up many other physics possibilities, for example, a high precision measurement of Japan's reactor antineutrinos, or using free neutrons to identify backgrounds in nucleon decay and atmospheric and accelerator neutrino studies.

*3.2. New Signals and Added Benefits For SK*

The ability to identify antineutrinos interacting via inverse beta decay gives SK the best chance of detecting SRNs for the first time ever. As shown by M. Nakahata, at the Neutrino 2008 Conference [9], if the Ando, Sato and Totani SRN model is assumed (with NNN05 flux revision), and SK is able to reduce the invisible muon background by neutron tagging, then with 10 years of data there would be a clear relic signal, especially in the visible energy range of 10 - 30 MeV. Figure 1a shows the energy spectrum as a function of visible energy in SK, with the assumed background levels, including neutron tagging, and a detection efficiency of 67%. The clear excess over background is seen at the lowest energies.

The addition of a Gd compound will allow SK to see antineutrinos coming from Japan's nuclear reactors, giving SK a second guaranteed new signal, and the ability to put better constraints on the solar oscillation parameters, using a single experiment. Figure 1b shows the results of a reactor antineutrino Monte Carlo study carried out at SK, using 15 nearby reactors, running for one year, with 4 different isotopes of reactor fuel, and the correct spectrum of each. The distance and direction vector to SK are included, as well as the average power of each reactor, giving a very conservative flux of ~2,800 events per year. In figure 1b the ratio of the oscillated spectrum to the unoscillated spectrum is shown, assuming $\Delta m_{12}^2 = 8.0 \times 10^{-5}$ and $\sin^2(2\theta_{12})=0.86$. A clear oscillation signature can be seen, so with real data and full statistical and systematic errors, an oscillation analysis can be carried out and a very competitive limit on $\Delta m_{12}^2$ placed, within 5 years.

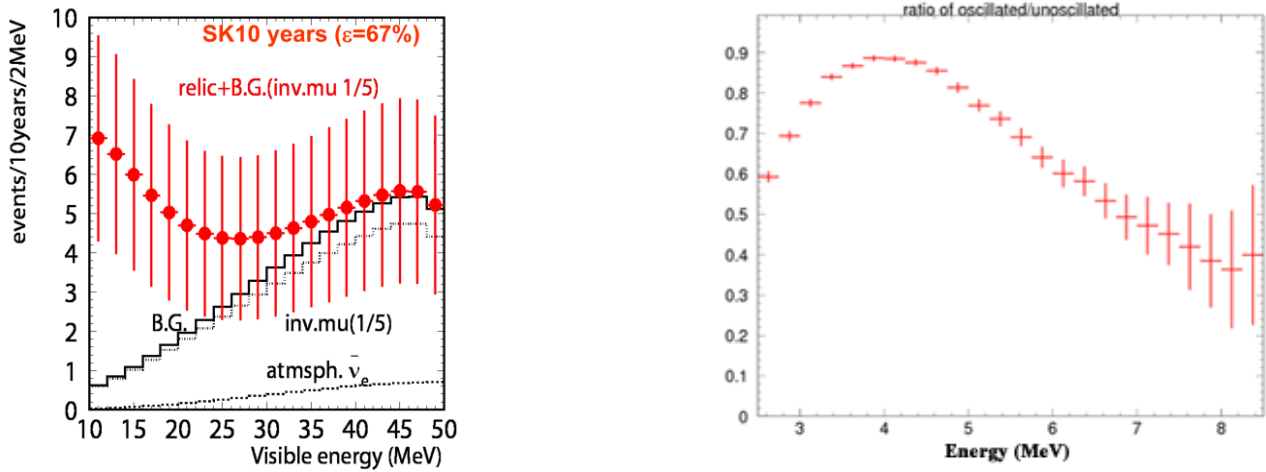

Fig. 1 (a) Expected SRN spectrum, assuming Ando, Sato and Totani relic model and an invisible muon rate reduction of 80% by neutron tagging; (b) Ratio of oscillated to unoscillated reactor spectrum from SK Monte Carlo study.

Aside from the two new signals described above, the ability to efficiently identify free neutrons in the SK tank will benefit many of the existing analyses. The ability to identify inverse beta events originating from a galactic supernova would allow these events to be separated from the remaining reactions, leading to up to 2 times better pointing accuracy to the supernova. This would also give the ability for very early warning of the supernova, detection of a faint supernova, or studying the burst with late time observations. Along these same lines, the solar antineutrino flux limit can also be improved. At higher neutrino energies, identifying (anti)neutrino events will separate these from other important events, allowing for a better study of hadronic final states in atmospheric and accelerator analyses. Even neutrino-neutron elastic scattering will become a new significant neutral-current channel in accelerator studies. Tagging the free neutrons can also aid in background reduction for proton decay analyses, and aid in the study of bound-nucleon decay in which a neutron is ejected [6].

## 4. Research and Development

### 4.1. First Gd Deployment in SK

To first investigate whether or not the 8 MeV gamma cascade of the free neutron capture on Gd could be seen within the SK tank, in 2007 an Am/Be source placed inside a BGO crystal, surrounded by a gadolinium trichloride ($GdCl_3$) solution, was deployed into the SK tank [10]. The alpha particles coming from the source captured on $^9$Be atoms giving carbons in an excited state and free neutrons. When the carbon de-excites, the produced 4.4 MeV gammas were seen as long and large time pulses in the BGO, while the free neutrons made their way through the BGO into the Gd solution, eventually capturing on Gd nuclei, giving the gamma cascade seen by the SK photomultiplier tubes (PMTs). The results of this study were published in [10], and showed that indeed free neutron captures on Gd nuclei could be seen inside the SK tank, and with the expected temporal and spatial separation from the initial BGO signals. This was the first time Gd had been successfully deployed, and the neutron capture signal seen, in a large water Cherenkov detector.

Since the time of this first Gd deployment into the SK tank, the best candidate Gd compound to dope a water Cherenkov detector with has changed from $GdCl_3$ to gadolinium sulfate ($Gd_2(SO_4)_3$) (motivation explained below). Because of the leading candidate Gd compound change, the same source and BGO crystal, this time surrounded by $Gd_2(SO_4)_3$, were deployed into the SK tank again in late 2010. The results were very comparable to the first deployment, with the signal clearly seen, and coming within the expected timing and distance from the BGO trigger. Figure 2a shows the timing of events from neutron capture on Gd, relative to the BGO trigger time. The average capture time is ~17.1 µs, while the large majority of neutrons captured within 35 µs. Figure 2b shows the vertex position, relative to the center of the source, and shows the large majority of captures occurring within 200 cm, giving a nice coincidence signal in the case of inverse beta decay.

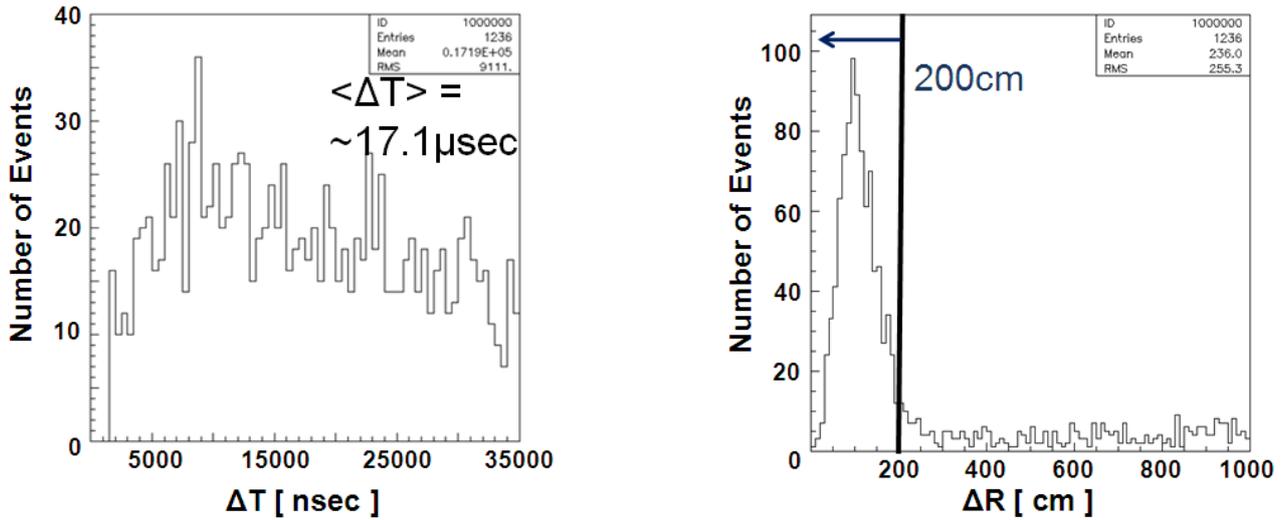

Fig. 2(a) Time distribution of free neutron capture events on Gd, relative to the BGO trigger, with the deployment of $Gd_2(SO_4)_3$ into SK; (b) Vertex distribution of the same neutron capture events, relative to center of source.

*4.2. Pushing the Gd Front*

As already mentioned above, although there is study going on in the US and Japan, the forefront of research and development for a Gd doped water Cherenkov detector is now being pushed at a brand new dedicated facility, called EGADS, located in the Kamioka Mine, in Japan. The EAGDS site is located just across the mine road from the SK detector, providing similar background rates, both cosmic and natural. EGADS has been built as a proof-of-principle experiment, planned to show that adding a Gd compound to a water Cherenkov detector is not only safe for detector components, but will also provide the required physics benefits. Between this facility and a few smaller labs in the US and Japan, every challenge faced in making this new method a reality is being addressed. The cumulative knowledge has been assembled at the EGADS facility, resulting in an experimental hall with everything needed to run and maintain a water Cherenkov detector doped with Gd.

There are many components to the new experimental facility, all of which are completely necessary to make the Gd idea work as it should, including the Gd compound itself. The main vessel is a 200 ton stainless steel (same metal as SK) tank, which will soon be fitted with 240 50 cm Hamamatsu PMTs (same as SK inner detector PMTs), all connected to a fully functioning, mock SK data acquisition (DAQ) system. Having a fully functioning DAQ will allow the study of existing and introduced backgrounds. Some of the PMTs will be enclosed in the same fiberglass and acrylic pressure vessels as those now installed in SK, while others will be un-housed. To keep the water circulating and actively clean it up, while not removing the Gd from solution, EGADS has the first working "selective band pass" water filtration system, a completely necessary part. To make sure the water system is doing its job in keeping the water transparent a special water transparency measurement device was developed at UCI, and a replica then built inside the EGADS hall. This enables real time measurements of water both inside the 200 ton tank, and also water coming out of different points in the water system. The last two major components are the Gd pre-treatment system and the Gd recovery system. The pre-treatment system was built to dissolve the $Gd_2(SO_4)_3$, in concentrated batches, clean up any uranium (U) which may have accompanied the $Gd_2(SO_4)_3$ from the manufacturer, and then inject the concentrated solution into the 200 ton tank. Once tests are completed, rather than releasing the Gd into the environment, two Gd recovery systems have been developed. These components, and the challenges they were built to overcome, will be described in the following sub-sections.

*4.3. Best Gd Compound Candidate*

Studies have been ongoing with three candidate Gd salt compounds; $GdCl_3$, $Gd_2(SO_4)_3$ and gadolinium nitrate ($Gd(NO_3)_3$), with $Gd_2(SO_4)_3$ being the clear choice to undergo a full scale trial at the EGADS site. There are a few requirements which a Gd compound must meet in order to become a good candidate for a full scale test. First the compound must be water soluble, and it should not be too difficult to dissolve large amounts, around 100 tons for a 50 kton detector. The above three candidates can all be dissolved fairly easily, with the $GdCl_3$ and $Gd(NO_3)_3$ only needing stirring to fully dissolve, while the $Gd_2(SO_4)_3$, can be forced into solution with the addition of a small amount of sulfuric acid, about 380 ml of acid for 28 kg of $Gd_2(SO_4)_3$ in 14 tons of water (test done at EGADS). So, this solubility requirement does not immediately rule out any of these three compounds.

Next, if the compound is to be used inside a very large water Cherenkov detector, such as SK, it must be safe for the detectors components and tank, and it should be a non-toxic chemical, so it may be easily put into existing detectors. While none of the above salts are toxic, they do have very different corrosive properties, due to the different anions of the molecule. The nitrate and the sulfate turn out to be non-corrosive, and do not seem to affect detectors components much, if at all. An extensive soak test study was carried out in Japan, with each of the 31 different materials inside the SK detector being soaked in both pure water and $Gd_2(SO_4)_3$ solution. Only the rubber used for the PMT pressure housing showed any difference between the pure water and the $Gd_2(SO_4)_3$ solution, with further studies showing the rubber is unaffected by the Gd solution, if the temperature is kept below $15^{\circ}$C, which it is in the case of SK. However, the chloride is corrosive, so for this reason the $GdCl_3$ has been ruled out as a full scale test candidate.

Last the Gd compound solution must maintain a high level of optical transparency, so that light can be seen from the opposite side of the detector. Although the $GdCl_3$ has fairly good optical properties, it has been ruled out by the above. $Gd(NO_3)_3$ is opaque in the UV region of the electromagnetic (EM) spectrum, for wavelengths less than 350 nm, and this unfortunately is where a large fraction of Cherenkov light is detected in SK PMTs, so it cannot be a good full scale test candidate either. This leaves $Gd_2(SO_4)_3$, and it turns out to have good transparency in the UV and optical regions of the EM spectrum, leaving it as the best candidate for the EGADS test facility. This is in fact the choice that has been made by the EGADS group, and the remaining studies described in this paper have been done using, or under the assumption that, $Gd_2(SO_4)_3$ will be used. One last exceptionally nice quality about this compound is today it costs only 5 US dollars per kilogram, meaning it would only be about 500,000 US dollars to dope SK [11], a price which is on the scale of such a major experiment.

*4.4. Gd Pre-treatment and Removal Systems*

To most easily and efficiently dissolve the Gd compound, and to be sure that it has all dissolved before injection into a detector, a pre-treatment system has been developed. The pre-treatment system at EGADS consists of a 15 ton tank with a large stirrer inserted into the center, UV lighting and mechanical pre-filters for bacteria removal, and a specially selected resin, which does an initial U removal without extracting the Gd. This pre-treatment system will be used to dissolve $Gd_2(SO_4)_3$, in highly concentrated batches, into ultrapure water inside the 15 ton tank. This highly concentrated solution will then circulate around the pre-treatment system as the special resin extracts U. Once the resin has extracted all the U it possibly can, the concentrated solution will be injected into the larger, 200 ton tank, where the full scale test will be taking place.

As mentioned before the Gd compound must not only be dissolved into the water, but it must also be extracted from the solution once the test is over, so no Gd is released into the environment. There are currently two methods which have been proven successful in recovering the Gd from solution, but only at the small scale level. The first uses ion exchange resin which is stored in large tanks, and is the first system built and ready at the EGADS site. The Gd solution will simply pass through these large tanks, while the ionized Gd is captured by the resin, and once saturated, the resin then disposed of properly. This method has been tested with the system at the EGADS site, giving Gd concentration reduction, from ~1,000 parts per million, to less than 0.5 parts per billion, well below the required limit. Unfortunately, this method of resin removal will only work for small scale experiments, as far too much resin would be required to extract Gd from a SK size detector. Thus, the second method under investigation at UCI uses a specially designed water system, to "drop" the Gd out of solution, by adding sodium hydroxide (NaOH), creating insoluble Gd hydroxide ($Gd(OH)_3$). It then passes the water and $Gd(OH)_3$ through a reusable filter press, where the Gd is then collected. This method is far more complex than the resin removal method, but it does not require the disposal and replacement of the large amounts of resin, and will only require NaOH and filter cleaning. Although concentrations of Gd coming out of the filter press are not quite as low as the resin removal system at this point, few parts per billion level, this method is being perfected at UCI, and eventually at EGADS, and is believed to be the preferred method of Gd removal for a large detector such as SK.

*4.5. Water Quality: Cleaning*

As the transparency of the Gd solution will have to be of the highest quality possible, it will have to be actively filtered at all times, just as SK is now. One of the main cleaning components of ultrapure water systems is de-ionizing (DI) tanks, which trap ions within them. If the Gd solution is passed through this type of water system, the DI tanks will absorb the Gd, until they are saturated, at which point they will no longer remove any ions at all. This will lower Gd concentrations tremendously, and once the DI resin saturates, the water will stop being cleaned to its full potential and optical quality will eventually drop, even with less Gd in solution. For this reason a new prototype water system was developed at UCI, using a method of "selective band pass" filtration, in which two paths are created through the water system based upon molecular size, one stream going through DI tanks and the other bypassing them. After proof that this system works at UCI, an industrial scale replica was built in the EGADS hall.

The first fork for Gd solution in the water system occurs at a nano-filter set, mechanical filters which reject ions and molecules larger than a certain pore size. The pore size of these filters is chosen to allow only things smaller than Gd through, giving a stream of water with a very low concentration of Gd, less than 0.15% the original concentration, in the prototype system at UCI. This low Gd concentration product stream goes into another identical nano-filter set, giving a second product stream with a Gd concentration less than 0.006% the original Gd concentration. This product stream is then sent through reverse osmosis (RO) filters and then DI tanks, giving tolerable levels of Gd loss. The ultrapure water coming from the DI tanks is then reunified with the two nano-filter rejected Gd streams, all of which is then sent through some mechanical filters, with pore sizes just big enough to let the Gd pass, and injected back into the main tank. Figure 3 shows a schematic drawing of the EGADS water system [11], lines showing the path of the Gd from the nano-filter reject labeled as "concentrated Gd NF reject lines", where it then goes through a pump, membrane degassifier, a filter and then into the main tank. This system was installed and has been recirculating water inside the 200 ton tank at the EGADS site, since January 2011. With the help of a water transparency measurement device (explained next) to measure the water, in less than 6 months the water system was able to bring pure water in the 200 ton tank up to a very high quality, comparable to the best seen in SK, with peak attenuation length > 120 meters at 405 nm, from very poor quality after having sat in the tank without recirculation for two weeks, which left peak attenuation lengths < 40 m at 405 nm. But even after only one month of recirculation, the transparency of the water in the 200 ton tank was already very good, with peak attenuation length > 100 m. The water system bringing the quality of the water up this high in the 200 ton tank, after such a short time, is a major feat for the Gd program.

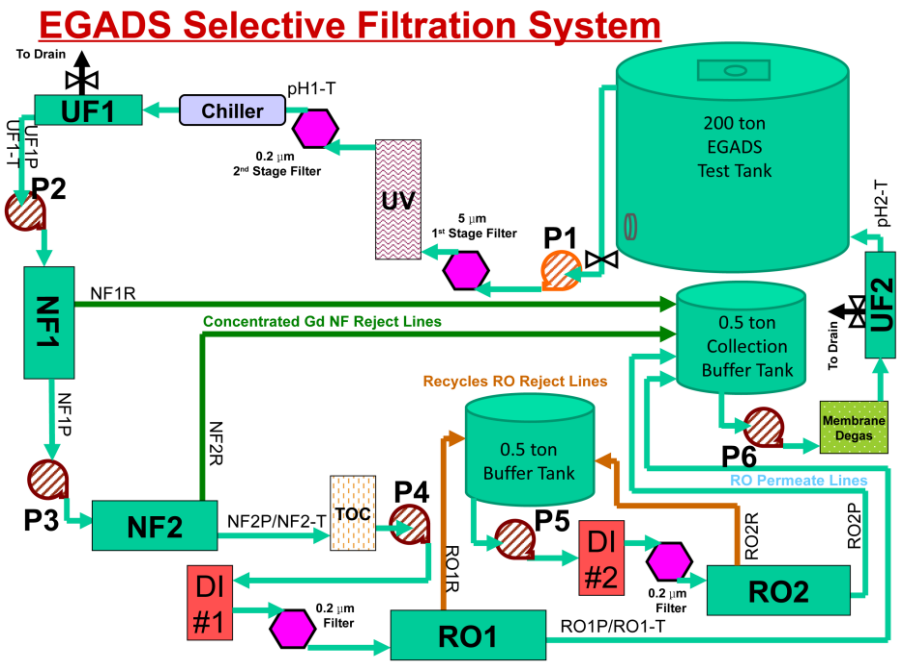

Fig. 3 EGADS "selective filtration" water system schematic view, the NF (nano) filters separate the Gd from the pure water stream which is sent through the DI tank, ensuring ultrapure Gd loaded water.

*4.6. Water Quality: Monitoring*

Measuring attenuation lengths of ultrapure water can be very difficult, as they may exceed 100 meters in the optical and UV region, and require long distances or very stable light monitoring to precisely measure them. Inside the SK tank this is achieved by using the large distances of the tank, as well as the ultra precise light counting by the PMTs, with a $4\pi$ solid angular view. When doing smaller development, large distances and ultra precise light monitoring are not as easily achieved. To overcome this problem a new device was built, first at UCI, called IDEAL (Irvine Device Evaluating Attenuation Length), followed by UDEAL (Underground Device Evaluating Attenuation Length), located at the EGADS facility. These two devices are basically identical, with the major difference being the height of the main pipe; IDEAL is 6.5 m in height, while UDEAL is extended to 8.6 m in height. On top of the pipe sits an array of 7 lasers, some laser pointers, some professional diodes and a nitrogen 337 nm laser, covering the range of Cherenkov light seen in SK. These lasers are pulsed, one at a time, into a beam splitter, one beam going into a monitoring 10 cm integrating sphere housing a UV enhanced silicon photodiode, and the other down through the main pipe of the device. The beam splitter can be steered by linear actuators, aiming the beam directly through the center of the main pipe and down into a 30 cm integrating sphere housing a similar silicon photodiode. The pulses at each integrating sphere are integrated, and the ratio of the integrated intensity taken as the measured value. This value is then measured for many pulses, for each laser, for a given height of the water column in the main pipe, giving an average ratio for each wavelength at a certain height. The height of the water column is varied from empty to full, and then to empty again, taking data points at discrete intervals along the way. A schematic of the devices can be seen in figure 4a.

A custom written, C++ program controls every hardware component of the devices, including automatic alignment of the lasers and all signal switching. It acquires signals from the photodiodes via a 12 bit PC oscilloscope, and computes each laser's attenuation length by fitting an exponential function to that laser's extinction curve. The inverse exponential coefficients of these fits give the attenuation lengths. Each time the main pipe is filled and drained while taking data defines a "data run", with each data run taking about 2 hours to complete. Figure 4b shows a run done at IDEAL, with the main pipe being filled with ~0.1% $Gd_2(SO_4)_3$ solution. Although the peak attenuation lengths of this run are greater than 80 m, the device is able to measure the attenuation length to ~1% statistical accuracy, with systematic errors still under study, but expected to be less than a few percent. UDEAL will monitor the quality of pure water and Gd solution in the EAGDS 200 and 15 ton tanks, as well as both coming from different points in the water system. IDEAL will now be used to do extensive soak studies of various materials, in both pure water and Gd solution, looking at what these materials do to the transparency of the water in the Cherenkov radiation region of the EM spectrum. One material in particular is a special submarine epoxy, which may be used to seal the leak present in the SK tank. This leak must be sealed if Gd is to ever be put inside SK, and so poses another challenge in making the GADZOOKS! idea a reality. Initial soak studies of the epoxy, with the use of correct filtration equipment, show the transparency can be held quite high, and with results from mechanical pressure tests also very good, this is a leading candidate method for fixing the leak.

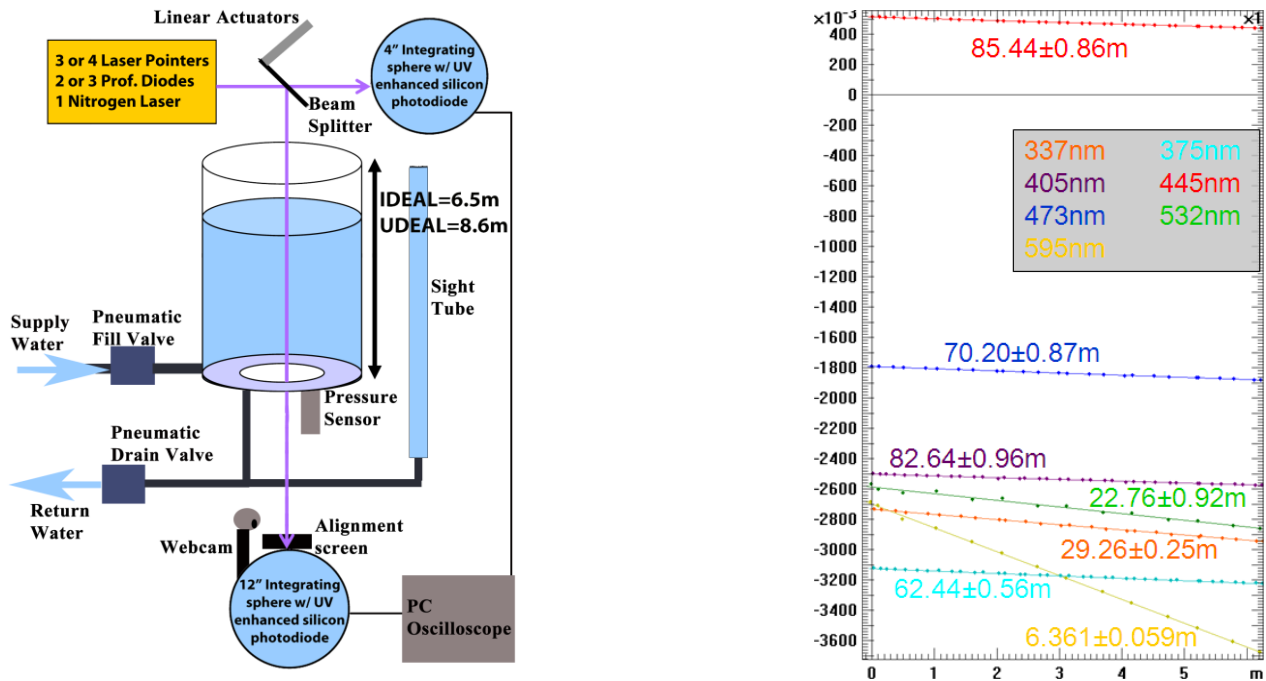

Fig. 4 (a) I/UDEAL schematic view (b) Gd solution run at IDEAL, with the measured attenuation lengths shown for each wavelength, the vertical scale is the log of the ratio of the integrated pulses, averaged over 300 pulses, from the two integrating spheres, while the horizontal gives the water column height in meters.

*4.7. New and Existing Background Levels*

The addition of $Gd_2(SO_4)_3$ into a water Cherenkov detector leads to the possibility of new backgrounds being introduced, in two different ways. Because the $Gd_2(SO_4)_3$ does not come 100% pure, anything accompanying it from the manufacturer will be put into the detector. Of course, while filtering may eventually take care of these contaminants, some may be the same size as the Gd ions, and so the water system would not be able to clean them up. One of these culprits could be U, with its radioactive decay looking just like any electron signal in the SK tank. To bring down U contamination a very special resin has been chosen, which very quickly saturates with Gd, while continuing to absorb U. Levels of U contamination were dropped to less than 1% the initial value, after the $Gd_2(SO_4)_3$ solution was passed through the resin at just 3 bed volumes. It is believed that the U levels can be brought down low enough to make little or no impact on physics studies. The second impact could come from the ability to efficiently see free neutrons in a large water Cherenkov detector for the first time. There have always been ambient neutrons in these detectors, but with the majority of them capturing on a proton and only producing a 2.2 MeV gamma, they have thus far escaped notice. Once they can be seen, the free neutrons will have to be identified and accounted for. One promising study showed that multiple scattering information, implemented through a "beta" parameter, as the SNO experiment did in their low energy paper [12], can discriminate between neutron capture events and true electron type events. Other initial studies show that ambient neutrons will not pose a problem, but these are ongoing and will have much more to say as the EGADS project proceeds.

**5. Current and Future Plans at EGADS**

With the completion of the water system in January 2011, EGADS is fully operational, but without any PMTs installed in the 200 ton tank. Initial running of the facility consisted of a pure water test, in which the 200 ton tank was filled with pure water, and then re-circulated through the water system for 6 months, all the time taking data runs with the UDEAL device. Initially the water system was turned off for two weeks, for installation, during which time UDEAL saw the transparency of the water degrade quickly and tremendously. Shortly after turning on the water system, within a week, the transparency of the water in the 200 ton tank was already of better quality than water inside SK when it was initially filled. This was a very promising first step in the EGADS project.

$Gd_2(SO_4)_3$ has now been successfully dissolved in the 15 ton tank, at a 0.2% concentration, and passed through the EGADS water system, with initial transparency measurements done. The water system is now being tuned, meaning flows and pressures adjusted, and different membranes tested in order to get the best Gd recovery rate in the nano-filter reject lines and the best water transparency. After circulation through the water system is optimal, the Gd solution will be injected into the 200 ton tank. After the solution in the 200 ton tank has been brought up to the highest possible quality, a more concentrated $Gd_2(SO_4)_3$ solution will be mixed in the 15 ton tank and then also injected into the 200 ton tank. The combined solution will recirculate around the water system and the 200 ton tank for a month or two, until all parameters look acceptable, at which point the tank will be drained, and a large scale Gd removal test done with the resin removal system. PMTs and DAQ will then be installed, the tank filled back up with pure water, and a new concentrated batch of $Gd_2(SO_4)_3$ solution made in the 15 ton tank. This will then be injected into the 200 ton tank, with PMTs in place, and the true full scale test will begin, hopefully early in 2012. These days are very exciting at the EGADS site, as each one poses new and interesting challenges and solutions, and all should stay tuned to see this new method of neutron detection in a large water Cherenkov detector develop.